\documentclass[sigconf,screen]{acmart}

\AtBeginDocument{%
  }

\usepackage{multirow}
\usepackage{booktabs}
\usepackage{float}
\usepackage{array}
\usepackage{amsmath,amsfonts} 
\usepackage{algorithmic}
\usepackage{graphicx}
\usepackage{textcomp}
\usepackage{xcolor}
\usepackage{listings}
\usepackage{tcolorbox}
\usepackage{bm}
\usepackage{enumitem}

\begin{document}

\title{\textsc{IssueGuard:} Real-Time Secret Leak Prevention Tool for GitHub Issue Reports}

\author{Md Nafiu Rahman}
\authornote{Equal contribution. Author order does not matter.}
\email{nafiu.rahman@bracu.ac.bd}
\affiliation{%
  \institution{Brac University}
  \city{Dhaka}
  \country{Bangladesh}
}

\author{Sadif Ahmed}
\authornotemark[1] 
\email{sadif.ahmed@bracu.ac.bd}
\affiliation{%
  \institution{Brac University}
  \city{Dhaka}
  \country{Bangladesh}
}

\author{Zahin Wahab}
\email{zahinwahab@gmail.com}
\affiliation{%
  \institution{The University of British Columbia}
  \city{Vancouver}
  \state{BC}
  \country{Canada}
}

\author{Gias Uddin}
\email{guddin@yorku.ca}
\affiliation{%
  \institution{York University}
  \city{Toronto} 
  \state{ON}
  \country{Canada}
}

\author{Rifat Shahriyar}
\email{rifat@cse.buet.ac.bd}
\affiliation{%
  \institution{Bangladesh University of Engineering and Technology}
  \city{Dhaka}
  \country{Bangladesh}
}

\renewcommand{\shortauthors}{Rahman et al.}

\begin{abstract}
GitHub and GitLab are widely used collaborative platforms whose issue-tracking systems contain large volumes of unstructured text, including logs, code snippets, and configuration examples. This creates a significant risk of accidental secret exposure, such as API keys and credentials, yet these platforms provide no mechanism to warn users before submission. We present \textsc{IssueGuard}, a tool for real-time detection and prevention of secret leaks in issue reports. Implemented as a Chrome extension, \textsc{IssueGuard} analyzes text as users type and combines regex-based candidate extraction with a fine-tuned CodeBERT model for contextual classification. This approach effectively separates real secrets from false positives and achieves an F1-score of 92.70\% on a benchmark dataset, outperforming traditional regex-based scanners. \textsc{IssueGuard} integrates directly into the web interface and continuously analyzes the issue editor, presenting clear visual warnings to help users avoid submitting sensitive data. The source code is publicly available at \href{https://github.com/disa-lab/IssueGuard}{https://github.com/disa-lab/IssueGuard}
, and a demonstration video is available at \href{https://youtu.be/kvbWA8rr9cU}{https://youtu.be/kvbWA8rr9cU}
.
\end{abstract}

\ccsdesc[500]{Security and privacy~Software and application security}
\ccsdesc[300]{Software and its engineering~Software post-development issues}

\keywords{Software Security, Secret Detection, Developer Tools, GitHub, Language Model, Chrome Extension}

\maketitle

\section{Introduction and Related Work}
Developers frequently handle sensitive credentials like API keys, OAuth tokens, and private encryption keys. Accidental leakage of these secrets presents a growing security risk, especially in modern distributed software systems \cite{secretsaboutsecretsincode}. The scale of this issue is significant; a 2022 report found over ten million new hardcoded secrets in public repositories \cite{secretsprawlreport}. While previous work has focused mainly on detecting secrets in source code and configuration files \cite{stealingAWScredentials,secretsaboutsecretsincode, meli2019bad, sinha2015detecting}, secrets also leak through unstructured channels such as issue reports, documentation, and developer discussions.

Existing research on secret detection has largely targeted source code \cite{meli2019bad, sinha2015detecting, saha2020secrets, feng2022automated, basak2023comparative}. Early approaches relying on regular expressions and entropy-based heuristics \cite{meli2019bad} often suffer from high false positive rates due to a lack of contextual understanding. While more recent machine learning-based approaches have moderately improved performance \cite{saha2020secrets, feng2022automated}, an effective detection model does not necessarily translate into a practical, preventative solution. Current platforms like GitHub and GitLab offer secret scanning, but these typically occur \textit{after} submission or within CI/CD pipelines \cite{trufflehog2, gitleaks2}. This reactive approach leaves developers vulnerable to exposing secrets in public logs before remediation can occur. Furthermore, conventional scanners like TruffleHog often generate high rates of false positives, leading to developer alert fatigue. Crucially, no existing tool provides real-time feedback within the issue-writing interface itself across these major platforms.

To bridge this gap, we present \textsc{IssueGuard}, a browser-based tool that brings real-time secret detection directly into the issue editors of GitHub and GitLab. Building on the contextual classification framework by Ahmed et al. \cite{ahmed2025secretbreachpreventionsoftware}, \textsc{IssueGuard} employs a two-stage pipeline: extracting candidates via regex and classifying them using a fine-tuned CodeBERT model. This approach significantly reduces false positives (achieving 92.70\% F1-score \cite{ahmed2025secretbreachpreventionsoftware}) leveraging contextual understanding and provides immediate, in-browser warnings as developers type, preventing leaks before they are posted. While our prior work introduced the foundational classification pipeline and evaluated its theoretical accuracy offline, \textsc{IssueGuard} focuses on the practical engineering required to operationalize that model. In addition to the browser extension, IssueGuard also supports GitHub/Gitlab CLI, enabling similar pre-submission checks in non-browser workflows. The main contributions of this work are:
\begin{enumerate}
    \item We introduce \textsc{IssueGuard}, a tool that integrates a transformer-based secret detection model into GitHub and GitLab (with Chrome extension and CLI support) to enable real-time, pre-submission leak prevention via server-side asynchronous processing, caching for latency optimizations, while maintaining usability through highlighting and client-side debouncing.
    \item We conduct a comprehensive usability study with 50 participants, demonstrating the tool's effectiveness and smooth integration into developer workflows.
\end{enumerate}

The rest of the paper is organized as follows. Section~\ref{sec:methodology} describes the system architecture. Section~\ref{sec:evaluation} assesses model effectiveness, real-time latency, and usability and Section~\ref{sec:conclusion} concludes the paper.

\section{Methodology}
\label{sec:methodology}
\begin{figure}
\centering
\includegraphics[width=0.9
\linewidth]{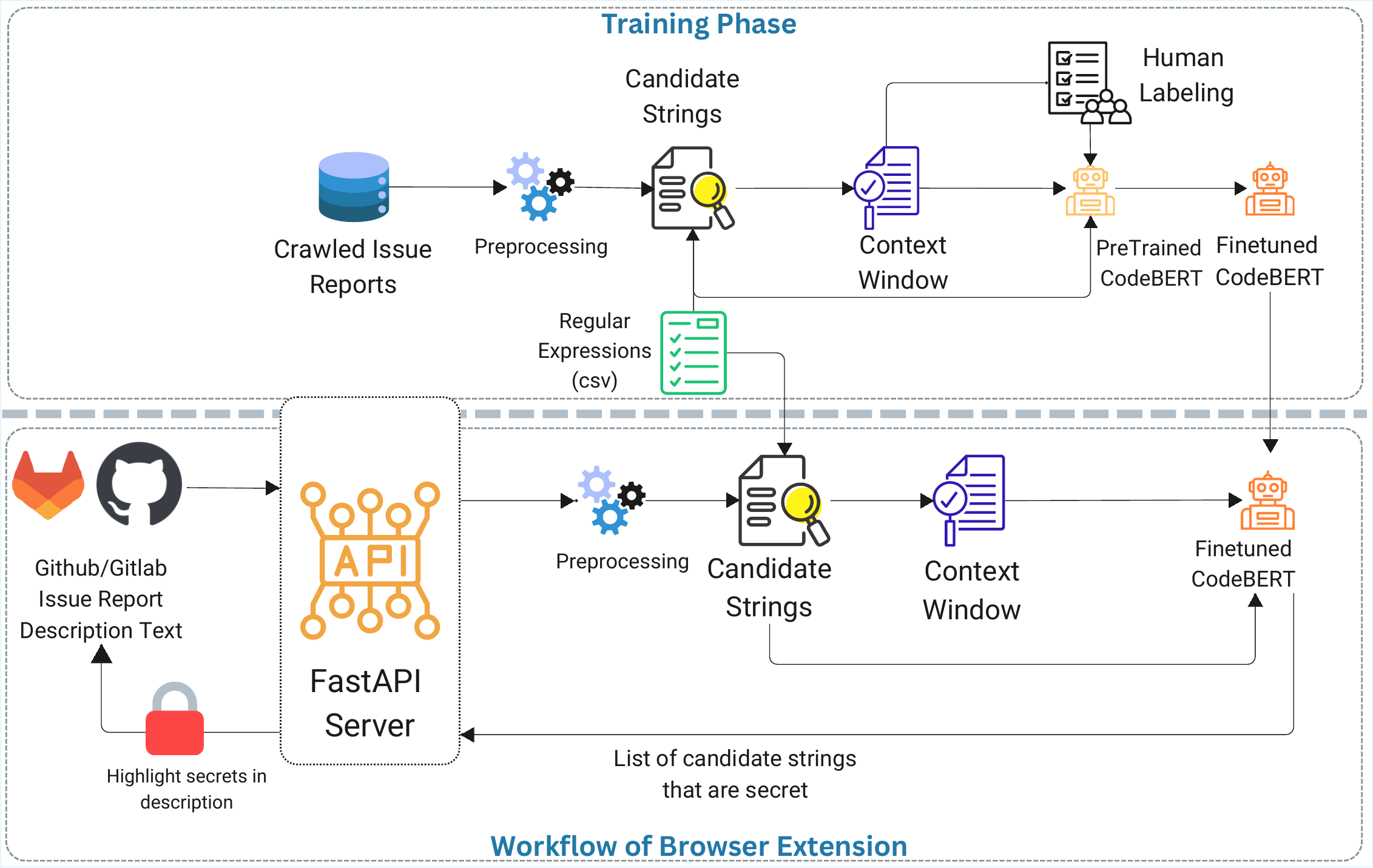}
\caption{Workflow of \textsc{IssueGuard}}
\label{fig:issueguard}
\end{figure}

In this section, we outline design and development of \textsc{IssueGuard} whose workflow is mentioned in Figure \ref{fig:issueguard}.

\subsection{Core Detection Engine}
\textsc{IssueGuard} is powered by the classification pipeline introduced in our prior work \cite{ahmed2025secretbreachpreventionsoftware}.
This engine serves as the backend for our tool.
We chose this pipeline as its underlying model, a fine-tuned CodeBERT, was shown to be effective at distinguishing real secrets from false positives in the environment of issue reports.
The pipeline’s model was trained on a benchmark dataset introduced in our prior work \cite{ahmed2025secretbreachpreventionsoftware}.
The dataset was constructed by collecting issue reports from GitHub repositories using keyword-based filtering (e.g., “key,” “token”), followed by applying 761 regular expressions to extract candidate secret strings.
These candidates were annotated as either \texttt{Secret} or \texttt{Non-sensitive} (e.g., placeholders, redacted values, or dummy keys), resulting in 54,148 labeled instances, including 5,881 true secrets, split into 75\% training, 10\% validation, and 15\% test sets.
For semantic feature extraction, the engine employs the \texttt{CodeBERT-base} model \cite{feng2020codebert}, a bimodal pre-trained model for programming and natural languages chosen for its performance on mixed-language tasks and fast inference speed.
A 200-character context window surrounding each candidate string is extracted from the issue report and concatenated with the candidate itself before being encoded by CodeBERT.
The resulting embedding is passed to a connected classification head trained using cross-entropy loss to predict the binary \texttt{Secret}/\texttt{Non-sensitive} label.
The fine-tuned model is then saved and integrated as the classification engine.

\subsection{\textsc{IssueGuard} System Architecture and Workflow}
\begin{figure}[h]
\centering
\includegraphics[width=0.9\linewidth]{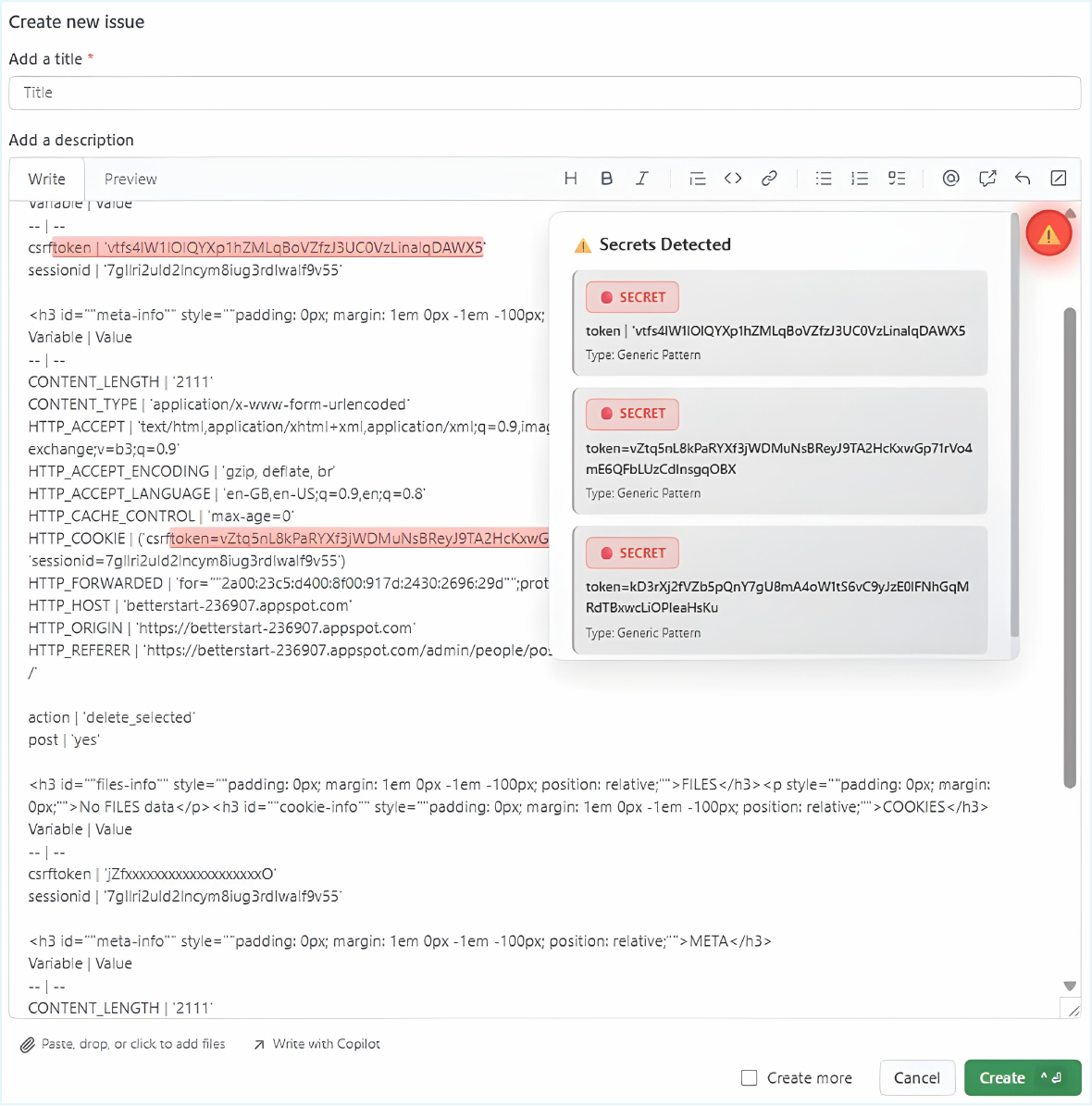}
\caption{Demonstration \textsc{IssueGuard}. As the user types, the extension sends text to the backend. Real secrets are highlighted in red, while regex-based positives such as placeholders are ignored.}
\label{fig:tool-demo}
\end{figure}

\textsc{IssueGuard} follows a client–server architecture consisting of a Google Chrome extension as the client and a FastAPI-based backend service as the server.
This design choice is driven by the goal to keep the browser extension responsive while performing expensive secret detection using a neural model.
By offloading model inference to the backend, the extension remains fast and unobtrusive during issue writing.
The tool is designed to integrate into developers’ existing workflows on GitHub and GitLab.
Figure \ref{fig:tool-demo} illustrates the workflow.
As developers write an issue report, \textsc{IssueGuard} performs background analysis and provides immediate feedback before the issue is submitted.
The detection pipeline proceeds follows:
\begin{enumerate}
\item \textbf{Text Monitoring:} The Chrome extension monitors paste events inside the issue description editor on GitHub/GitLab.
\item \textbf{Debounced Requests:} To avoid communication during continuous typing, the extension applies a debouncing strategy, pausing backend requests until the user stops typing.
\item \textbf{Deferred Transmission:} The current text is sent to the backend only after a short pause in user input.
\item \textbf{Candidate Extraction:} Upon receiving the text, the backend applies a set of 761 regular expressions to identify potential secret candidates.
\item \textbf{Contextual Classification:} Each candidate is paired with a 200-character context window and passed to the fine-tuned CodeBERT-based classifier.
\item \textbf{False Positive Filtering:} The classifier filters out benign strings such as placeholders, hashes, and redacted values.
\item \textbf{Result Transmission:} The backend returns confirmed secrets to the browser extension.
\item \textbf{User Feedback:} The extension highlights detected secrets in the editor and displays a tooltip warning, allowing users to correct the issue before submission.

Beyond the browser-based interface, IssueGuard also provides GitHub/Gitlab CLI support to accommodate developer workflows outside the web editor.
\end{enumerate}
Low latency is a requirement for interactive developer tools.
To achieve this, \textsc{IssueGuard} incorporates several engineering optimizations.
On the client side, debouncing reduces backend requests and limits network overhead.
On the server side, FastAPI is used for its asynchronous and non-blocking execution model, allowing the backend to handle concurrent requests efficiently.
Model inference is executed in a thread pool so that expensive computations do not block request handling.
When running on GPU-enabled systems, the backend uses mixed-precision (FP16) inference \cite{micikevicius2018mixed}, which reduces memory usage and improves throughput without affecting prediction quality.
In addition, a Least Recently Used (LRU) cache stores inference results.
This allows repeated analyses of similar text to be returned, further improving responsiveness during iterative editing. We evaluate the system’s performance by deploying the backend on a machine with an AMD Ryzen 3700G CPU, 16 GB RAM, and an NVIDIA RTX 3060 GPU.
The architecture supports local execution within environments, ensuring that issues with sensitive information remain under the user’s control, avoiding the privacy concerns associated with cloud-hosted security.

\section{Evaluation}
\label{sec:evaluation}
In this section, we evaluate the effectiveness of \textsc{IssueGuard}. We assess three key aspects: 1) the classification effectiveness of its underlying model compared to state-of-the-art LLMs, 2) the real-time performance and latency of the tool itself, and 3) its practical usability and developer acceptance, measured via a user study.

\subsection{Model Effectiveness}

As outlined in the methodology, \textsc{IssueGuard} integrates the CodeBERT-based classification pipeline from our prior work \cite{ahmed2025secretbreachpreventionsoftware}. While newer Large Language Models (LLMs) such as StarCoder \cite{li2023starcodersourceyou} and specialized encoders like StarPII  \cite{allal2023santacoderdontreachstars} have shown promise in code tasks and personal information detection, our evaluation confirms that fine-tuned CodeBERT remains the optimal choice for real-time secret detection in issue reports.

To validate this choice, we compared our model against several other models, including StarCoder (15B), StarPII, and RoBERTa-base. As shown in Table \ref{tab:model_comparison}, our fine-tuned CodeBERT model achieves an F1-score of 92.70\%, outperforming both the much larger StarCoder (89.01\%) and the PII-specialized StarPII (90.24\%). While StarPII shows high precision, it suffers from lower recall in the noisy context of issue descriptions. CodeBERT provides the best balance of precision and recall while remaining lightweight enough for low-latency inference.

\begin{table}[h]
\centering
\caption{Comparative analysis of detection models}
\label{tab:model_comparison}
\footnotesize
\setlength{\tabcolsep}{4pt}
\begin{tabular}{|l|c|c|c|}
\hline
\textbf{Model} & \textbf{Precision} & \textbf{Recall} & \textbf{F1-Score} \\
\hline
\textbf{IssueGuard (CodeBERT)} & \textbf{92.49\%} & \textbf{92.91\%} & \textbf{92.70\%} \\
RoBERTa-base & 91.10\% & 89.40\% & 90.24\% \\
StarPII (BigCode) & 93.05\% & 85.50\% & 89.11\% \\
StarCoder (15B) & 88.50\% & 89.50\% & 89.00\% \\
Regex-only & 6.80\% & 100.0\% & 12.80\% \\
\hline
\end{tabular}
\end{table}

Furthermore, the model's generalization was validated on a set of 178 real-world GitHub repositories randomly selected from popular public projects. As shown in Table \ref{tab:wildresult}, the model achieved a macro-average F1-score of 81.60\%, demonstrating a strong ability to generalize beyond its training data, adopted from \cite{ahmed2025secretbreachpreventionsoftware}).

\begin{table}[h]
\caption{Model performance on 178 real-world repositories}
\centering
\footnotesize
\setlength{\tabcolsep}{6pt}
\begin{tabular}{|c|c|c|c|c|}
\hline
\textbf{Class} & \textbf{Precision} & \textbf{Recall} & \textbf{F1-score} \\
\hline
\textbf{Secret} & 50.98\% & 86.67\% & 64.20\% \\
\textbf{Macro Avg.} & 75.35\% & 92.45\% & 81.60\%  \\
\hline
\end{tabular}
\label{tab:wildresult}
\end{table}

\subsection{Comparison with Standard Security Tools}

Existing secret detection tools such as TruffleHog and Gitleaks are industry standards for post-commit remediation, detecting secrets after they have entered source code or logs. However, using these repository-level scanners in real-time prevention workflows leads to alert fatigue. As shown in Table~\ref{tab:tool_comparison}, TruffleHog and Gitleaks achieve high recall (over 93\% and 90\%), making them well suited for security audits, but their regex-based approaches yield low precision (around 50\%), which would cause frequent false alerts in an interactive setting. In contrast, \textsc{IssueGuard} is designed for pre-submission use and operates directly on issue reports, an attack surface not covered by existing scanners. By leveraging semantic context from a fine-tuned CodeBERT model, it filters out common false positives such as placeholder values, achieving over 92'\% precision while maintaining high recall (F1-score: 92.70\%). Thus, while traditional scanners protect repositories post-commit, \textsc{IssueGuard} provides a high-precision, pre-commit defense that prevents secrets from leaving the developer’s local environment.

\begin{table}[h]
\centering
\caption{Performance comparison of IssueGuard with other secret scanners. }
\label{tab:tool_comparison}
\footnotesize
\setlength{\tabcolsep}{6pt}
\begin{tabular}{|l|c|c|c|}
\hline
\textbf{Tool} & \textbf{Precision} & \textbf{Recall} & \textbf{F1-Score} \\
\hline
\textbf{IssueGuard} & \textbf{92.49\%} & \textbf{92.91\%} & \textbf{92.70\%} \\
TruffleHog (v3.0) & 55.39\% & 93.94\% & 69.69\% \\
Gitleaks (v8.18) & 49.10\% & 90.20\% & 63.60\% \\
\hline
\end{tabular}
\end{table}

\subsection{Tool Performance and Latency}
Beyond model accuracy, a real-time tool must be \textit{fast}. As detailed in our methodology, \textsc{IssueGuard}'s architecture is highly optimized with asynchronous processing and caching. Latency was measured as the time from completing a typing event in the GitHub issue editor to the browser’s visual highlighting of detected secrets. To ensure a rigorous evaluation, these measurements were conducted using the test split defined in our dataset, ensuring a representative mix of short and long issue descriptions.

The results show an average end-to-end prediction time of 198 milliseconds (0.198 seconds). The core model inference step is extremely efficient, accounting for only 10.1 milliseconds (0.0101s) of this time. This demonstrates that the tool's architecture is well-suited for uninterrupted real-time application.

\subsection{User Study on Tool Usability}

To evaluate practical utility, we conducted a user study with 50 participants (25 developers, 12 testers, 6 team leads, and 7 managers). Participants were recruited via internal company mailing lists and university developer forums to ensure a diverse range of experience (0–10+ years). We specifically targeted standard software engineering roles rather than specialized security personnel, as accidental secret leaks are a pervasive issue among general developers rather than just security experts. While some participants had prior affiliations with the authors, none had previous exposure to the tool. To mitigate potential bias, all participants were provided with strictly standardized, identical instructions.

Participants were tasked with submitting 10 issue reports on GitHub, five using the \textsc{IssueGuard} extension and five manually, with instructions to identify potential secrets (e.g., AWS keys, private tokens) within the reports. We acknowledge that this setup inherently primes users to be cautious, which differs slightly from the casual nature of inadvertent real-world leaks. However, this allowed us to effectively measure the tool's utility as an active safety net. Despite being primed, 66\% of participants still inadvertently submitted overlooked secrets during the manual inspection phase;  \textsc{IssueGuard} successfully flagged these instances in the assisted phase. Following the task, we administered a survey based on Kitchenham and Pfleeger’s guidelines \cite{kitchenham2008personal}. Open-ended survey responses were analyzed using thematic coding. Two authors independently reviewed the responses, assigned initial codes, and then merged similar codes into broader themes through discussion until agreement was reached. The qualitative questions asked participants about their overall experience with IssueGuard, perceived strengths, limitations, and suggestions for improvement. The results, summarized in Table~\ref{tab:user_study}, were highly positive. In particular, participants reported significantly higher confidence when using \textsc{IssueGuard} compared to manual inspection(3.3/5), noting that the tool reduced the risk of overlooking subtle or partially redacted secrets. 

\begin{figure}[h]
    \centering
    \includegraphics[width=0.8\linewidth]{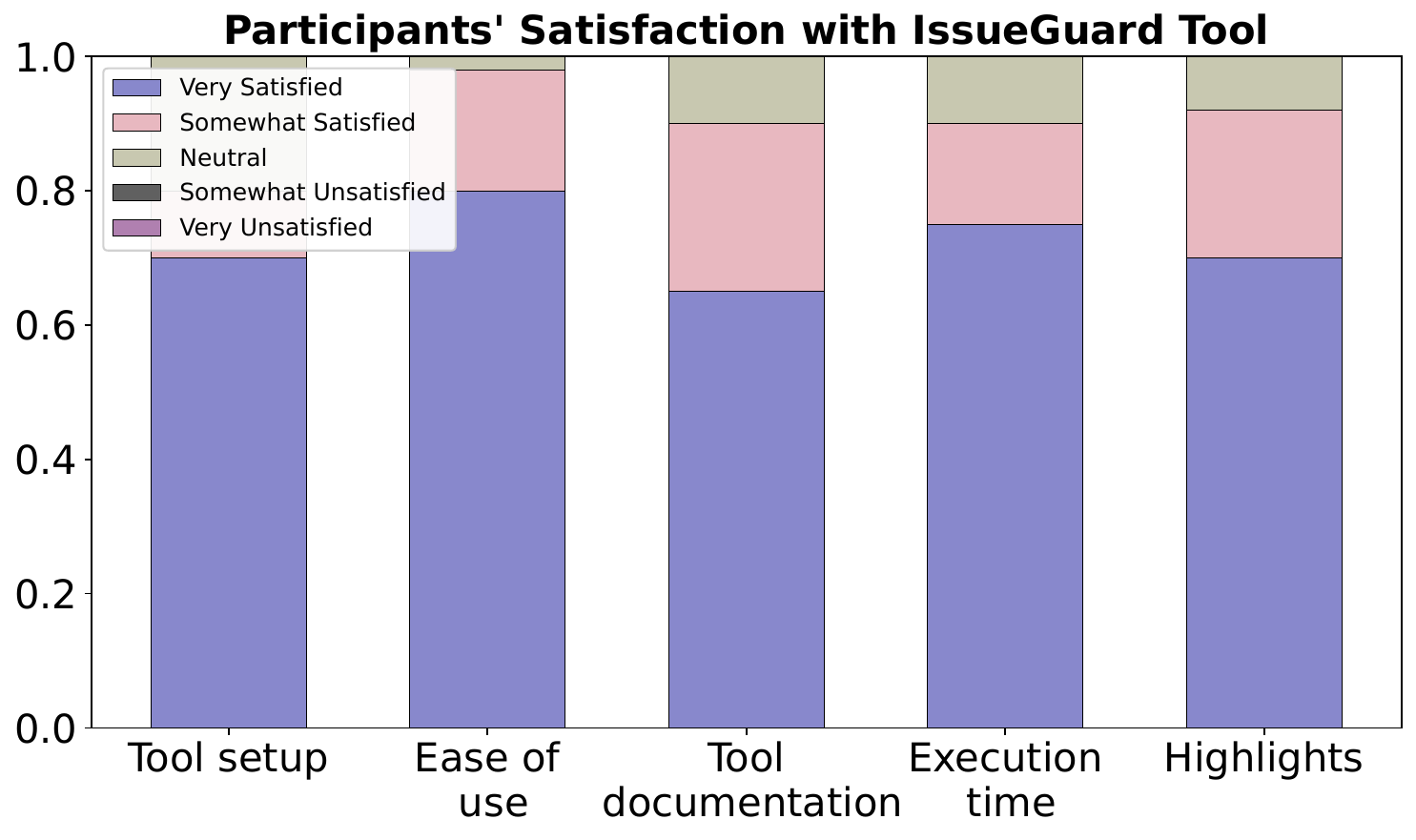}
    \caption{Participants' satisfaction with various aspects of
    the \textsc{IssueGuard} tool.}
    \label{fig:issueguard_satisfaction}
\end{figure}

Overall, 90\% of participants rated their confidence in the tool at 4 or higher (on a 5-point scale), while feature satisfaction was similarly strong, with 80\% of users reporting being “Very Satisfied” with ease of use. Qualitative feedback further highlighted the tool’s “native” feel, with users emphasizing that unobtrusive highlights reduced the cognitive load associated with manually searching for sensitive information.
Feature satisfaction (Figure \ref{fig:issueguard_satisfaction}) was similarly exceptional.
A vast majority reported being `Very Satisfied' with the tool setup (75\%), ease of use (80\%), execution time (78\%), and highlights (72\%). Even tool documentation, which is typically a lower-rated aspect in software tools, was highly praised, with 93\% of users expressing satisfaction. Qualitative feedback from the open-ended questions verified the quantitative data. Participants frequently noted that the presence of real-time feedback increased their confidence in submitting issues publicly, suggesting that IssueGuard not only detects leaks but also improves developers’ perceived security assurance.

\begin{table}[h]
\centering
\caption{User Study: Confidence in \textsc{IssueGuard}'s Detection}
\label{tab:user_study}
\footnotesize
\setlength{\tabcolsep}{6pt}
\begin{tabular}{|c|c|c|c|c|}
\hline
\textbf{Confidence Rating (1-5)} & \textbf{\# of Participants} & \textbf{Percentage} \\ 
\hline
5 (Very High Confidence) & 40 & 80.0\% \\
4 (High Confidence) & 5 & 10.0\% \\
3 (Moderate Confidence) & 4 & 8.0\% \\
2 (Low Confidence) & 1 & 2.0\% \\
1 (Very Low Confidence) & 0 & 0.0\% \\ 
\hline
\textbf{Total} & \textbf{50} & \textbf{100.0\%} \\ 
\hline
\end{tabular}
\end{table}

\section{Limitations and Threats to Validity}
\label{sec:threats}
While \textsc{IssueGuard} shows strong performance, it has limitations. Its two-stage pipeline uses regular expressions for initial extraction, prioritizing precision and low latency over full coverage, so heavily obfuscated secrets or unusual formats outside the 761 predefined patterns may be missed. False negatives can also arise from truncated contexts or ambiguous placeholder credentials \cite{ahmed2025secretbreachpreventionsoftware}. Evaluating on our own benchmark may introduce dataset bias, but no public benchmarks exist for secret leaks in unstructured issue reports. Additionally, the client-server architecture, even when local, raises adoption and privacy concerns for enterprises. Future work will focus on a lightweight, event-driven execution model.

\section{Conclusion and Future Scope}
\label{sec:conclusion}

In this work, we present \textsc{IssueGuard}, a practical tool to prevent accidental secret leaks in issue reports. It combines a CodeBERT-based contextual classifier with a low-latency Chrome extension for real-time detection during issue submission. Evaluation shows high accuracy with real-time responsiveness, demonstrating effective, developer-facing preventative security. We also extend IssueGuard with a GitHub/Gitlab CLI for terminal workflows, with plans to support more platforms and IDEs as future work.

\bibliographystyle{ACM-Reference-Format}
\bibliography{references}

\end{document}